# Evaluation of the Extension of the Cerebral Blood Flow and its Main Parameters

**Alexander Gersten**
Department of Physics,
and Unit of Biomedical Engineering,
and Zlotowski Center for Neuroscience,
Ben-Gurion University of the Negev
Beer-Sheva 84105, Israel
E-mail: gersten@bgumail.bgu.ac.il

### Abstract

Among the major factors controlling the cerebral blood flow (CBF) - cerebral perfusion pressure (CPP), arterial partial pressure of oxygen ($P_{aO_2}$), cerebral metabolism, arterial partial pressure of carbon dioxide ($P_{aCO_2}$), and cardiac output (CO), the effect of $P_{aCO_2}$ is peculiar in being independent of autoregulatory CBF mechanisms and it allows to explore the full range of the CBF. We have developed a simple physical model, and have derived a simple four parameter formula, relating the CBF to $P_{aCO_2}$. The parameters are: $B_{max}$, the maximal CBF, $B_{min}$ the minimal CBF, $p_0$ the value of $P_{aCO_2}$ at the average CBF and the parameter A, the slope at this point. The parameters can be extracted in an easy way, directly from the experimental data. With this model five experimental data sets of human, rats, baboons and dogs were well fitted. The same type of parametrization was also used successfully for fitting experimental data of $P_{aO_2}$ of dogs. We have also looked on the dependence of the $P_{aCO_2}$ parameters on other factors and were able to evaluate their dependence on the mean arterial blood pressure.

# 1. **Introduction**

In this work it was found that breathing may have dramatic effects on the brain blood flow. This was already known long time ago to Chinese, Indians and Tibetans. Here we shall add some simple mathematical models which allow a quantitative description of cerebral blood flow.

In recent years there was a considerable progress made in utilizing measurements of the regional cerebral blood flow (rCBF) in order to study brain functioning (Knezevic et al.,1988, Angerson et al.,1989, Ginsberg and Scheinberg, 1991, Costa and Ell, 1991, Howard, 1992). But it seems that the physical and mathematical aspects of the global cerebral blood flow (CBF), or average rCBF, were not sufficiently explored. Our main interest is of using physical principles (Hobbie, 1988) and physical and mathematical reasoning and means in order to describe in a simple way the main features of CBF.

The human brain consists of about 2% of the adult body weight, but consumes (at rest) about 15% of the cardiac output (CO) and about 20% of the body's oxygen demand (Sokoloff, 1989, Guyton, 1991).

Glucose is the main source of cellular energy through its oxidation (Sokoloff, 1989, Moser, 1988). The cerebral glucose utilization is almost directly proportional to the CBF , (Harper, 1989, McCulloch, 1988, Harper and McCulloch, 1985). The CBF can be influenced by abnormal glucose levels, is increased during hypoglycemia (Horinaka et al., 1997) and decreased during hyperglycemia Duckrow, 1995).

Normal mean CBF is approximately 50-55 ml/100g/min, but declines with the age (above the age of about 30), in a rate of approximately 58.5-0.24×age ml/100g/min (Maximilian and Brawanski, 1988, Hagstadius and Risberg, 1983, see also Yamamoto et al., 1980, for other details).

The cardiac output can be increased many times (up to about tenfold) during very heavy exercise or work (see Appendix A), But only part of the cardiac output increase can be accommodated by the brain blood vessels because of autoregulatory mechanisms and of the vessels limited capacitance, which is influenced by their elasticity, limited space of the cranium and the presence of the cerebrospinal fluid (CSF).

There exist autoregulatory mechanisms which maintain the CBF approximately constant for cerebral perfusion pressure (CPP) over an approximate range of 60-160 mm Hg (Harper, 1989, McCulloch, 1988, Aaslid *et al.,* 1989, Ursino, 1991). Outside this autoregulatory range the CBF may decrease (CPP<60 mm Hg) as in the case of hypotonia (Sokoloff, 1989) or increase (CPP>160 mm Hg) as in the case of high hypertension (Guyton, 1991). But again, the above statements are valid only for normal functioning.

For some abnormal functioning the autoregulatory mechanisms may break down, for example if $Pa_{CO_2} > 70$ mm Hg (Harper, 1989, Harper, 1966).

The CBF is also influenced by the value of cerebral tissue $P_{O_2}$, which normal range is 35-40 mmHg. A decrease below approximately 30 mmHg will cause the increase of CBF (Guyton, 1991), for details see sec. 5.

The main parameter influencing the CBF is the arterial $Pa_{CO_2}$. About 70% increase (or even less) in arterial $Pa_{CO_2}$ may double the blood flow (normal value of $Pa_{CO_2}$ is about 40 mmHg.) (Sokoloff, 1989, Guyton, 1991). A.M. Harper (Harper, 1989) gave an interesting and vivid description of the above situation:

"However, in the welter of PET scanners, NMRs and SPECTS, with physicists, isotope chemists, computer experts and mathematicians producing reams of data and results accurate to the nth decimal place from pixel sizes which are shrinking by the week, it is easy to forget that a few deep breaths from the patient could lower his arterial carbon dioxide tension by 2 mm Hg (0.27 kPa) and reduce his cerebral blood flow by about 5 per cent. Were this to go unnoticed, the efforts put into our measurements will have been in vain in respect of interpretation of the data for a clinical purpose."

The CBF is very sensitive to $P_{aCO_2}$ and it is our aim to demonstrate in a simple physical model that important information about CBF capacitance can be obtained by considering only the dependence of CBF on $P_{aCO_2}$. Slowing down the breathing rate, without enhancing the airflow (Fried and Grimaldi, 1993, Fried, 1987, Fried 1990, Fried 1999, Timmons and Ley, 1994), or holding the breath, can increase $P_{aCO_2}$. It is plausible that this is one of the essences of yoga pranayama (Bernard, 1960, Iyengar, 1981, Joshi, 1983, Kuvalayananda, 1983, Lysebeth, 1979, Rama, 1986, 1988, Shantikumar, 1987, Shrikrishna, 1996] and of Tibetan six yogas of Naropa (Evans-Wentz, 1958, Garma, 1963, Mullin, 1991,1996, 1997, Yeshe, 1998). It seems that biofeedback training of breathing (Fried and Grimaldi, 1993, Timmons and Ley, 1994), or methods advocated in yoga, may become important for treating health problems.

## 2. Cerebral blood flow and arterial partial pressure of carbon dioxide

Let us first examine the existing data of the CBF as a function of arterial $CO_2$. Let us start with the older data. In (Longobardo et al., 1987) the data of (Salazar and Knowls, 1964) are presented as:

$$\dot{Q}_{brain} = \left(0.04 \frac{L}{mmHg \cdot min}\right) Pa_{CO_2} - 0.93 \ \frac{L}{min} \ , \qquad (2.1)$$

without giving the validity range. From (Guyton, 1991) we can deduce that the relation is approximately linear in the range

$$37 \text{ mmHg} < Pa_{CO_2} < 60 \text{ mm Hg} \ . \qquad (2.2)$$

For a brain weight of about 1.4 kg, eq.(1) can be presented in units more commonly used:

$$\dot{Q}_{brain} = \left(2.9 \frac{mL}{mmHg \cdot 100g \cdot min}\right) Pa_{CO_2} - 66 \cdot mL/100g/min \ , \qquad (2.3)$$

for 100g of brain tissue. Out of the range of the inequality (2.2) the CBF becomes much less sensitive to the changes of $Pa_{CO_2}$. Equation (2.3) with the restriction (2.2) will be used here as the first approximation. Of course one can expect individual changes. In hyperventilation (Fried and Grimaldi, 1993) one goes below the lower limit of eq.(2.2), therefore one needs a more accurate and meaningful treatment. Below we derive a more accurate description of the data.

Inspection of experimental data, especially the more accurate ones on animals, like ones done with rhesus monkeys (Reivich, 1964), or with rats (Siesjo and Ingvar, 1986, Sage et al., 1981) led us to conclude that the CBF (which will be denoted later as B) is limited between two values. This we can interpret in the following way: the upper limit $B_{max}$ corresponds to maximal dilation of the blood vessels and the lower (non-negative) limit $B_{min}$ to the maximal constriction of the vessels. We can incorporate these requirements using the following assumptions:

$$\frac{dB}{dp} = AF\left(\frac{B - B_{min}}{B_{max} - B_{min}}\right) = AF(z) \geq 0, \qquad z = \frac{B - B_{min}}{B_{max} - B_{min}} \qquad (2.4)$$

$$0 \leq z \leq 1, \quad F(0) = F(1) = F'(0) = F'(1) = 0, \quad 0 \leq B_{min} \leq B \leq B_{max},$$

where A is a constant and the condition F(0)=0 corresponds to the requirement that the constriction is maximal, F(1)=0 correspond to maximal dilation. Another physical boundary constraint can be formulated, for the derivatives F'(z) of F(z), as follows: F'(1)=F'(0)=0, meaning that the approach to the limits is not abrupt but smooth. A simple choice satisfying these requirements is:

$$F(z) = \sin^2(\pi z), \qquad (2.5)$$

which also allows to integrate analytically eq. (2.4). This is so as from eq.(2.4):

$$dB = (B_{max} - B_{min})dz, \qquad (2.6)$$

and one can easily check that

$$\int \frac{dz}{\sin^2(\pi z)} = -\frac{1}{\pi}ctg(\pi z) + \text{Cons} \tan t \, . \qquad (2.7)$$

From Eq. (2.7)

$$\int_{z_1}^{z_2} \frac{dz'}{\sin^2(\pi z')} = -\frac{1}{\pi}ctg(\pi z_2) + \frac{1}{\pi}ctg(\pi z_1) = \frac{1}{\pi}\frac{\sin \pi(z_2 - z_1)}{\sin \pi z_2 \sin \pi z_1} \, . \qquad (2.8)$$

and in particular

$$\int_{\frac{1}{2}}^{z} \frac{dz'}{\sin^2(\pi z')} = -\frac{1}{\pi}ctg(\pi z). \qquad (2.9)$$

Eqs. (2.4) and (2.5) can be now converted to

$$\frac{dz}{\sin^2(\pi z)} = \frac{Adp}{(B_{max} - B_{min})}. \qquad (2.10)$$

Integrating Eq. (2.10), using Eq. (2.8), we obtain

$$\frac{1}{\pi}\frac{\sin\left[\frac{\pi}{\Delta B}(B - B_r)\right]}{\sin\left[\frac{\pi}{\Delta B}(B - B_{min})\right]\sin\left[\frac{\pi}{\Delta B}(B_r - B_{min})\right]} = \frac{A(p - p_r)}{\Delta B}, \qquad (2.11)$$

where the index r is associated with a reference point $p_r$ to which corresponds CBF of $B_r$, and

$$\Delta B = B_{max} - B_{min}. \qquad (2.12)$$

One can invert Eq. (2.11) if for the reference point for CBF the value half way between the extremes is being taken. A simple presentation is obtained if in addition to Eq. (2.12) one introduces the constants:

$$B_0 = B(z = \tfrac{1}{2}) = \tfrac{1}{2}(B_{max} + B_{min}); \quad ; \quad \left(\frac{dB}{dp}\right)_{B=B_0} = A, \quad p_0 = p(z = \tfrac{1}{2}). \tag{2.13}$$

Remembering that $p \equiv Pa_{CO_2}$, from Eqs. (2.11), (2.9) and the relation:

$$\text{arc ctg}(\pi z) + \text{arc tg}(\pi z) = \tfrac{\pi}{2},$$

we can finally obtain:

$$B(p) = B_0 + \frac{\Delta B}{\pi} \cdot \text{arctg}\left(\frac{\pi(p - p_0)A}{\Delta B}\right), \quad p > 0 \tag{2.14}$$

where $B_0$ is the half way value of the CBF (z=1/2) between the maximal and minimal B, $p_0$ is the value of p corresponding to $B_0$, and A is the slope at this point. All these parameters can be estimated easily, directly from the data.

Approximate values of $B_{min}$, $B_{max}$, $p_0$ and A, for the data of Fig. 2.1 based on (Wyngaarden, 1992) (the circles are not the actual data, but they were extracted from the curve fitting the data) are given below:

$$\begin{aligned} B_{min} &= 3 \text{ cc}/100\text{g}/\text{min}, \ B_{max} = 114 \text{ cc}/100\text{g}/\text{min}, \\ A &= 1.9 \text{ cc}/100\text{g}/\text{min}/\text{mmHg}, \ p_0 = 43 \text{ mmHg}. \end{aligned} \tag{2.15}$$

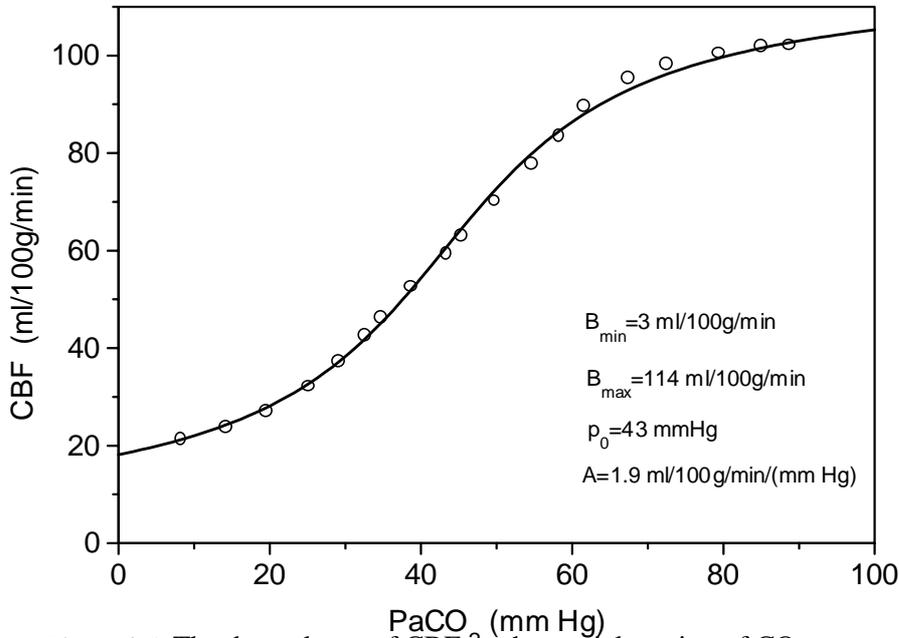

*Figure 2.1* The dependence of CBF on the partial tension of $CO_2$.
The fit is according to eqs. (2.7) and (2.8). The data (circles) are based on
(Wyngaarden, 1992). No indication is given which animal was experimented.

This result should be considered only as a typical example, not as a universal one. Please note the large difference in the slopes (2.15 against 1.9) in eqs. (2.3) and (2.15).

Approximate values of $B_{min}$, $B_{max}$, $p_0$ and A, for the data (solid curve) of Fig. 2.2 based on (Reivich, 1964) are given below:

$$B_{min} = 10 \text{ cc}/100g/\min, \quad B_{max} = 100 \text{ cc}/100g/\min,$$
$$A = 1.9 \text{ cc}/100g/\min/mmHg, \quad p_0 = 47 \text{ mmHg}. \tag{2.15a}$$

The doted line is a best fit to the data with

$$B(p) = 20.9 + \frac{92.8}{1 + 10570 \exp[-5.251 \, \log(p)]}. \tag{2.15b}$$

The data in Fig.2.2 are based on experiments with 8 animals. They are scattered to large extent, therefore one can suspect that there are large individual differences between the animals. Indeed the data of each animal separately are more regular. In Fig. 2.3 the data of the 5th animal are shown. Approximate values of $B_{min}$, $B_{max}$, $p_0$ and A, for the data (dotted curve) of Fig. 2.3 based on (Reivich, 1964) are given below:

$$B_{min} = 0 \text{ cc}/100g/\min, \quad B_{max} = 120 \text{ cc}/100g/\min,$$
$$A = 1.5 \text{ cc}/100g/\min/mmHg, \quad p_0 = 47 \text{ mmHg}. \tag{2.15c}$$

For rats the data of (Sage et al., 1981) are fitted with:

$$B_{min} = 75.38 \pm 4.19 \text{ cc}/100g/\min, \quad B_{max} = 480.4 \pm 11.4 \text{ cc}/100g/\min,$$
$$A = 18.82 \pm 1.62 \text{ cc}/100g/\min/mmHg, \quad p_0 = 36.38 \pm .78 \text{ mmHg}. \tag{2.16}$$

and are displayed in fig. 2.4. Inspecting eq. (2.7) we find that

$$B_{min} = B(-\infty), \qquad B_{max} = B(\infty), \tag{2.17}$$

i.e. $B_{min}$ is in the non-physical (non-physiological) region of p, far away from the physical region. Therefore the results should not depend significantly on its value. The maximum constriction should be estimated from the CBF at the border of the physical region:

$$B(0) = B_0 - \frac{\Delta B}{\pi} \cdot \text{arctg}\left(\frac{\pi p_0 A}{\Delta B}\right). \tag{2.18}$$

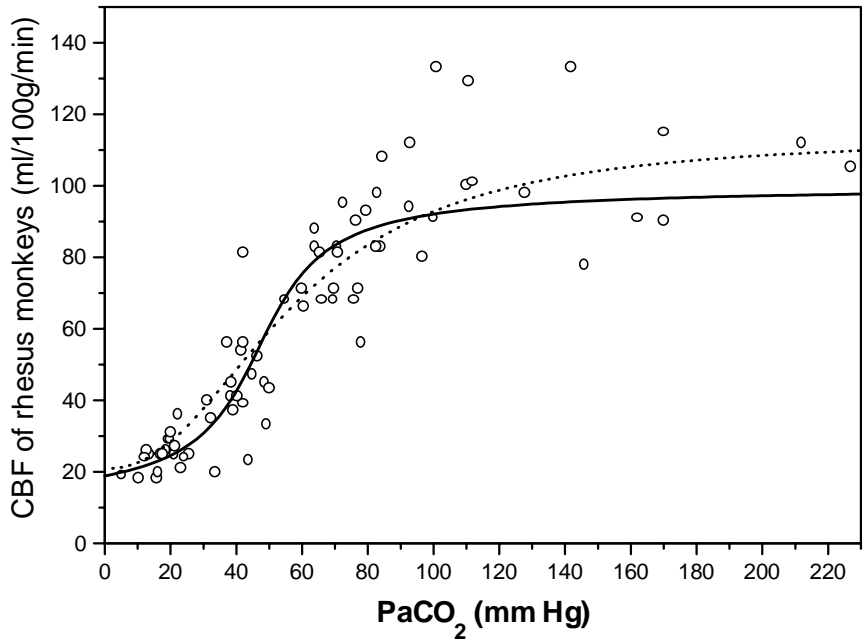

*Figure 2.2* The dependence of CBF of rhesus monkeys on the partial tension of $CO_2$. The continuous curve is the approximate fit to eq. (2.7) for the data of (Reivich, 1964). The dotted line is the fit with eq. (2.15b).

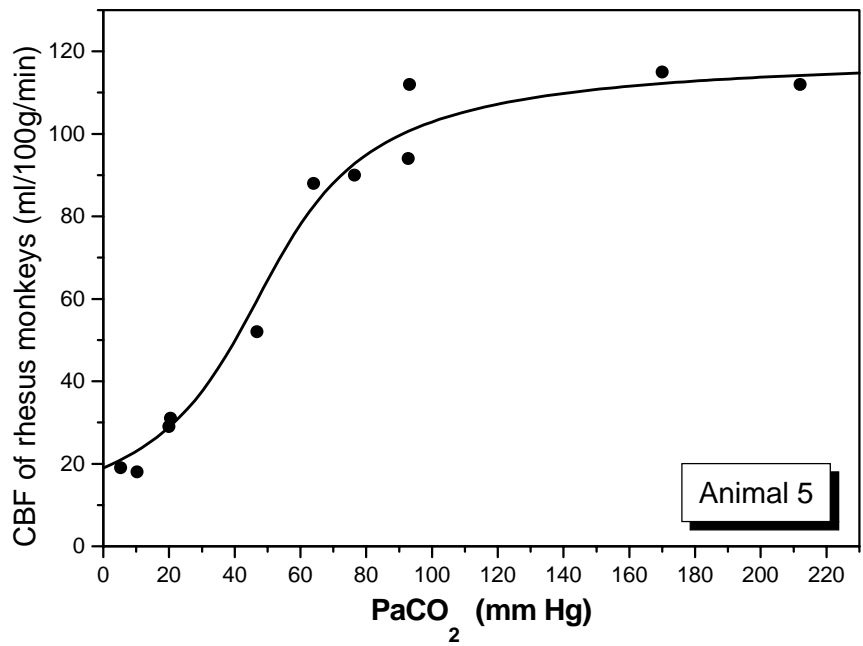

*Figure 2.3* The dependence of CBF of rhesus monkeys on the partial tension of $CO_2$. The continuous curve is the approximate fit eq. (2.15.c) for the data of (Reivich, 1964).

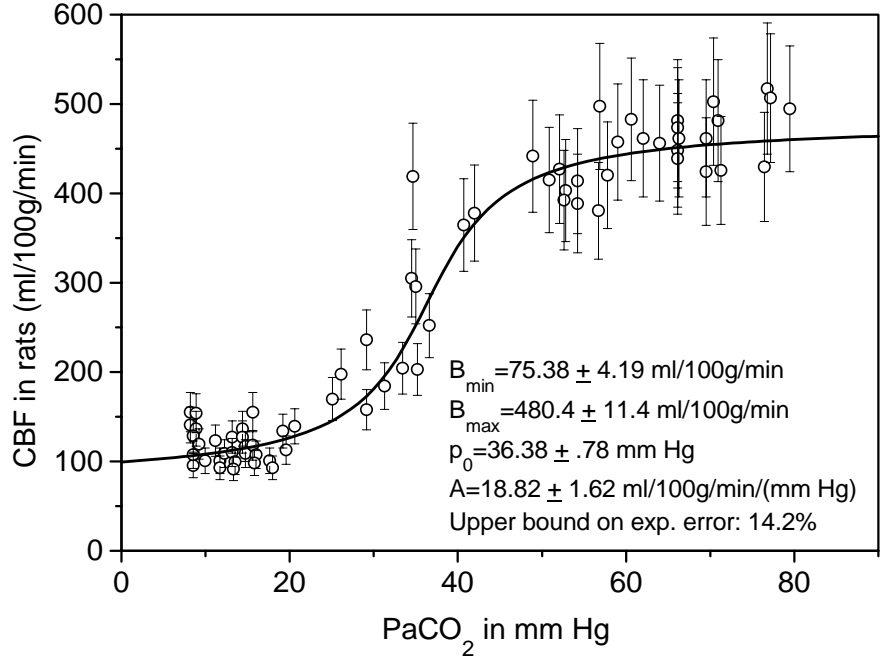

*Figure 2.4* The dependence of CBF of rats on the partial tension of $CO_2$. The continuous curve is the best fit to eq. (2.7) for the data of (Sage et al., 1981)]. The error bars were added to the data according to the procedure outlined in sec. 3.

## 3. Error estimation

While performing the best fit (the least square fit) to experimental data, the chi squared

$$\chi^2 = \sum_{n=1}^{N} \frac{(t_n - e_n)^2}{(\Delta e_n)^2}, \qquad (3.1)$$

is being minimized with respect to the searched parameters ($B_{min}$, $B_{max}$, $p_0$, A). Above, in eq. (3.1), $e_n$ are the experimental data, $\Delta e_n$ their errors (standard deviations), N the total number of experimental points, and $t_n$ are the theoretical predictions (from eq. (2.7), in our case, depending on the parameters $B_{min}$, $B_{max}$, $p_0$, A ). The data that we are using do not have the experimental errors evaluated. Therefore we are enforced to use some assumptions, which may lead to reasonable results. Our first assumption is that the errors are proportional to the measured results (i.e., there is a fixed percentage error C):

$$\Delta e_n = C \cdot e_n . \qquad (3.2)$$

Let $\chi_0^2$ be the minimal value of eq. (3.1). If the theory is exact, the expectation value for the chi squared is

$$\chi_0^2 = N, \qquad (3.3)$$

otherwise

$$\chi_0^2 > N. \qquad (3.4)$$

Substituting eq. (3.2) into eq. (3.1) and taking into account the two possibilities (3.3) or (3.4) we obtain:

$$C \le \sqrt{\frac{1}{N}\sum_{n=1}^{N}\frac{(t_n - e_n)^2}{(e_n)^2}} = C_{UB}, \qquad (3.5)$$

which gives us an upper bound for the percentage error. Throughout the paper we use in the figures (only when the error bars are not supplied) for error bars the values:

$$\Delta e_n = C_{UB} \cdot E_n. \qquad (3.6)$$

The error of a parameter is defined as the value of the change of this parameter from the best fit, which causes a change in the chi squared by one unit. This is the procedure, which will be used throughout the paper, for estimating the errors of the model parameters. Thus if near the local minimum at $P_0$ with respect to a parameter P, the chi squared behaves like

$$\chi^2 = \chi_0^2 + \frac{1}{2}(P - P_0)^2 \frac{\partial^2 \chi^2}{\partial P^2}\Big|_{\chi^2 = \chi_0^2}, \qquad (3.7)$$

the error $\Delta P$ of the parameter P can be evaluated from

$$\chi^2 - \chi_0^2 = 1 = \frac{1}{2}\Delta P^2 \frac{\partial^2 \chi^2}{\partial P^2}\Big|_{\chi^2 = \chi_0^2}. \qquad (3.8)$$

In fig. 3.1 we present a fit to data for which the errors (standard deviations) were carefully evaluated. There are there 10 experimental points of the CBF of baboons (Branch et al., 1991, Ewing et al., 1989, Ewing et al., 1990) as a function of partial arterial tension of $CO_2$. The resulting fit parameters are inside fig. 3.1. It is interesting to note the value of the chi squared 10.04 is very close to the expectation value 10 (data points). This give more confidence, not only for the evaluated standard deviations, but also for the theory.

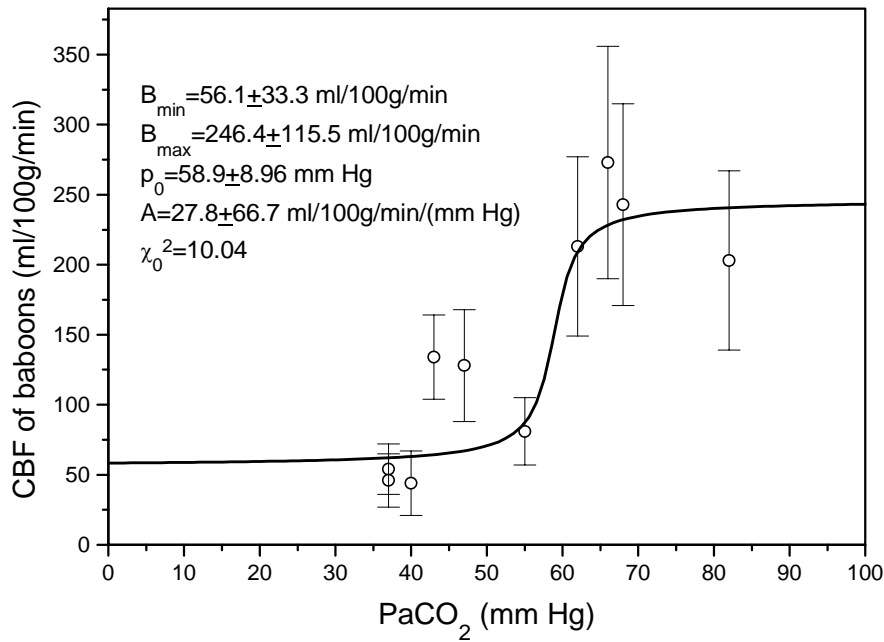

*Figure 3.1* The dependence of CBF of baboons on the partial tension of $CO_2$. The data with their errors (standard deviations) were taken from (Branch et al., 1991, Ewing et al., 1989, Ewing et al., 1990)

## 4. Modification of the Cerebrovascular Responses of $P_{aCO_2}$

It was emphasized in sec. 2 that the experimental data cited there were based on some measurements done on individuals and that the parameters may be dependent on other factors as well. In sec. 1, for example, the age factor was indicated. In experiments performed on dogs it was found that the cerebrovascular dilation induced by hypercapnia is attenuated by hypotension (Harper and Glass, 1965).

In this section we will discuss the dependence of the parameters given by the model of eq. (2.7) on the mean arterial blood pressure (MABP). There are some experimental data on dogs (Harper, 1989, McCulloch, 1988, Harper, 1966, Harper and Glass, 1965) which will allow us to do so. In figs. 3, 4 and 5 the data (circles) are presented for MABP=150, 100, and 50 mm HG. The error bars attached to them were computed according to the procedures of sec. 3.

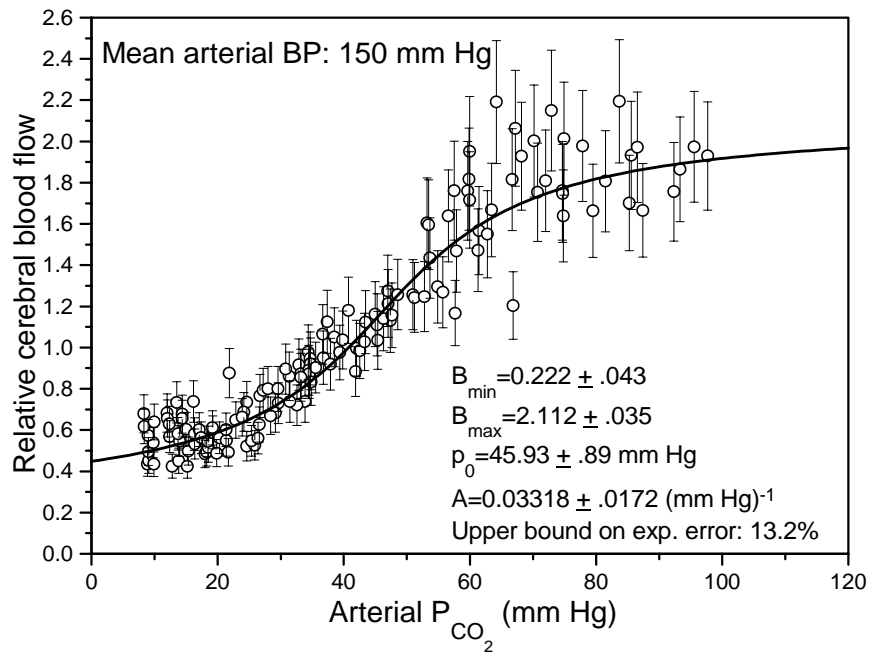

*Figure 4.1* The dependence of CBF of dogs on the partial tension of $CO_2$, relative to its value at 40 mm Hg, for mean arterial blood pressure of 150 mm Hg. The data are based on (Harper and Glass, 1965). The error bars were added to the data according to the procedure outlined in sec. 3.

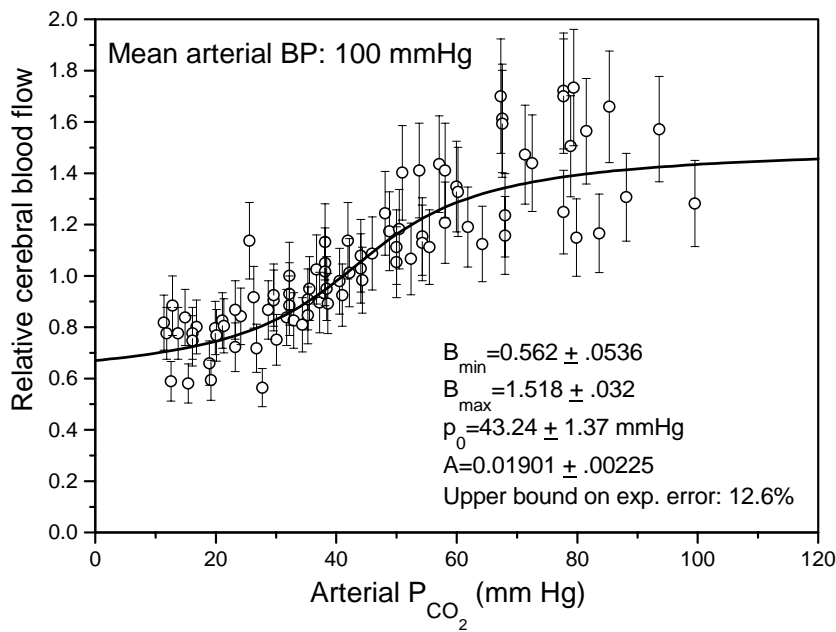

*Figure 4.2* The dependence of CBF of dogs on the partial tension of $CO_2$, relative to its value at 40 mm Hg, for mean arterial blood pressure of 100 mm Hg. The data are based on (Harper and Glass, 1965)]. The error bars were added to the data according to the procedure outlined in sec. 3.

For the 150 and 100 mm Hg data the chi squared had definite minimum. The results of the fits are inside the figures 4.1 and 4.2.

The fit to the 50 mm Hg data was more problematic, as the data (see fig. 4.3) are almost consistent with a straight line (equal to one) and there was no definite minimum. We had to impose some constraints on the parameters. It seemed to us as a reasonable assumption to assume that for AMBP=0, the CBF=0, and consequently $B_{min}=B_{max}=1$, and also A=0. For the low values of AMBP=50 mm Hg we expected the above parameters to converge to their AMBP=0 values. In the fit the most unstable was the A parameter. As we expected it to converge to zero for AMBP=0 we used as a constraint the demand that in the search it will reach its lowest value in a local minimum. This value of A was further kept unchanged. The results of that constrained search are given inside fig. 4.3.

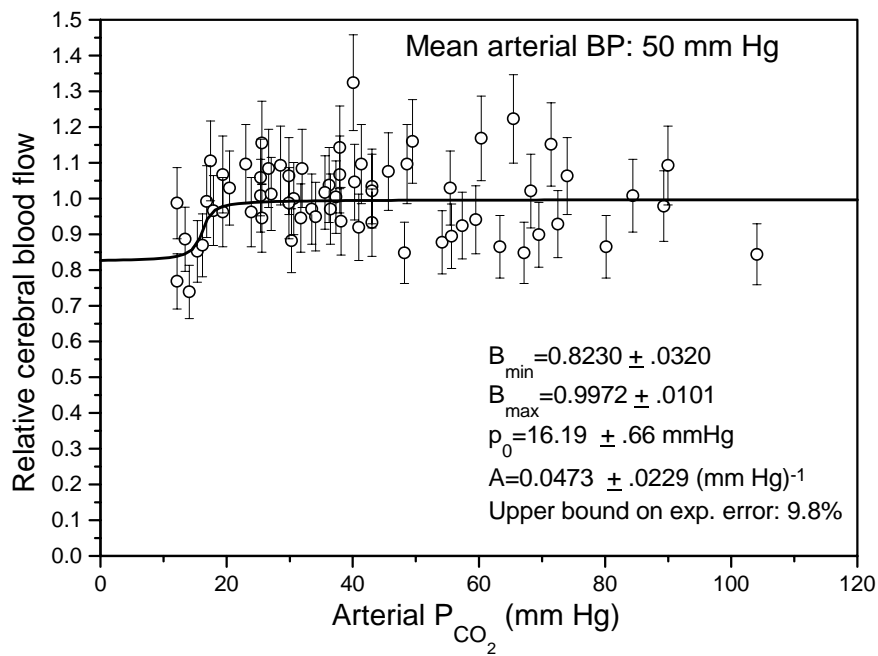

*Figure 4.3* The dependence of CBF of dogs on the partial tension of $CO_2$, relative to its value at 40 mm Hg, for mean arterial blood pressure of 50 mm Hg. The data are based on (Harper and Glass, 1965). The error bars were added to the data according to the procedure outlined in sec. 3.

The parameters, which were obtained in the 3 searches are displayed in figs. 4.4-4.7.

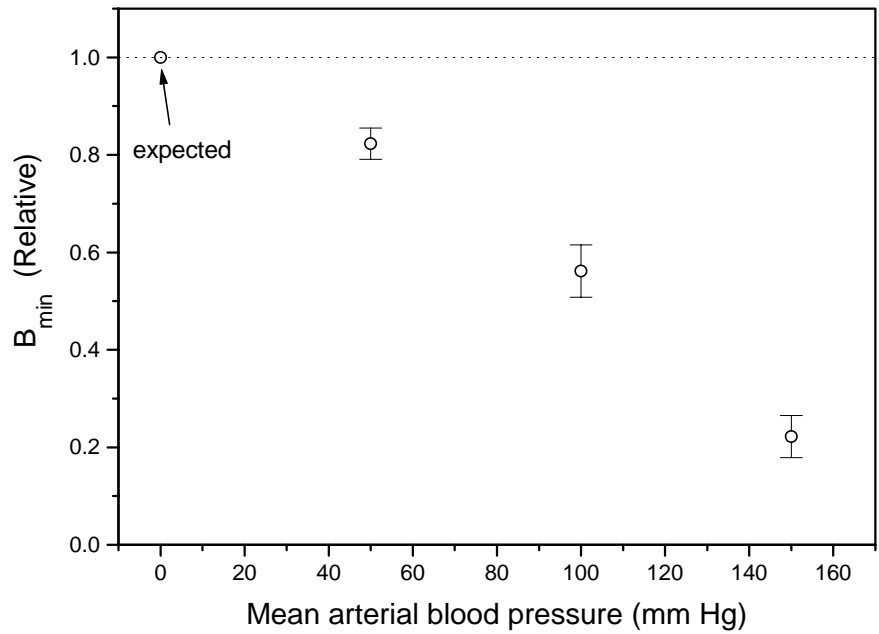

*Figure 4.4* The best fit values of the parameter $B_{min}$ the cases presented in figures 4.1, 4.2, and 4.3 against the mean arterial blood pressure.

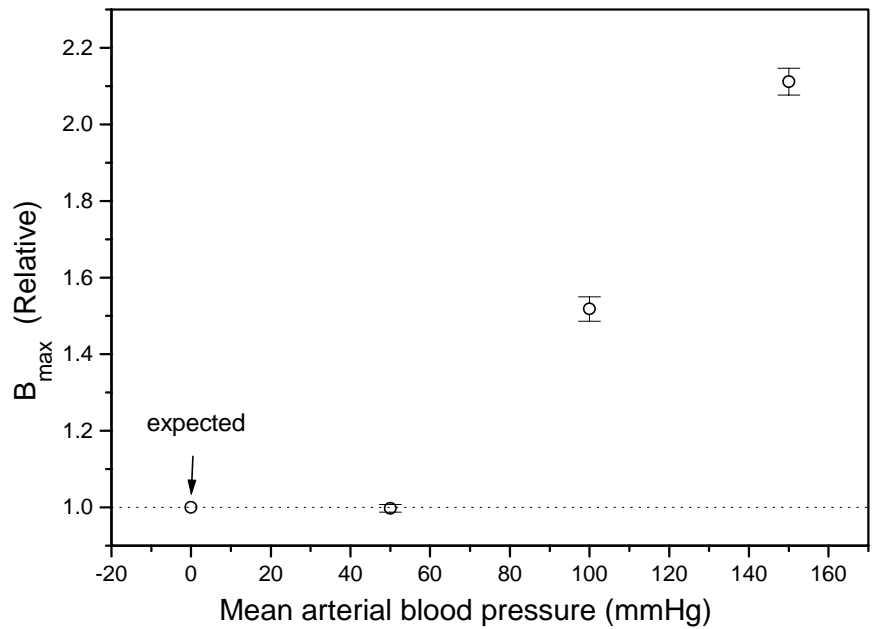

*Figure 4.5* The best fit values of the parameter $B_{max}$ for the cases presented in figures 4.1, 4.2, and 4.3 against the mean arterial blood pressure.

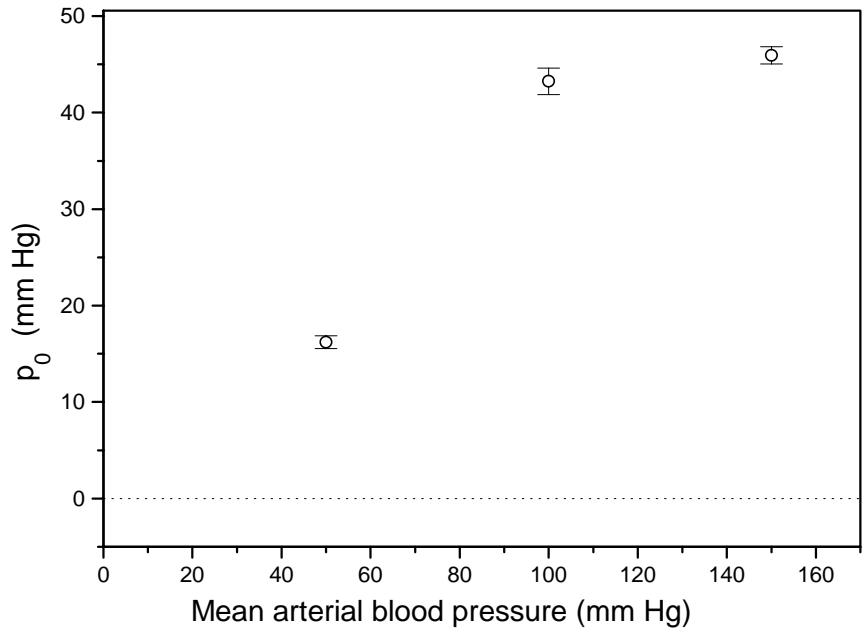

*Figure 4.6* The best fit values of the parameter $p_0$ for the cases presented in figures 4.1, 4.2, and 4.3 against the mean arterial blood pressure.

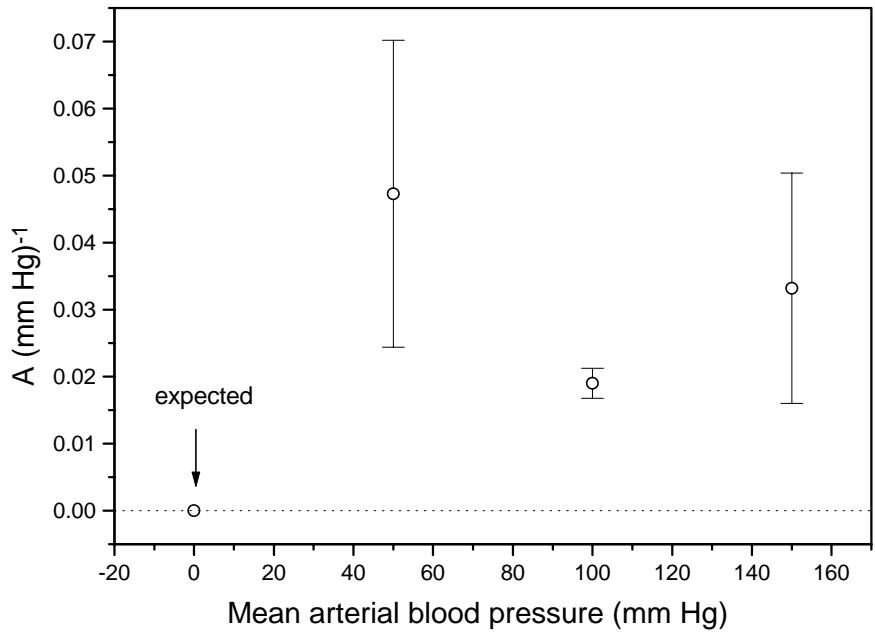

*Figure 4.7* The best fit values of the parameter A for the cases presented in figures 4.1, 4.2, and 4.3 against the mean arterial blood pressure.

## 5. Cerebral blood flow and arterial partial pressure of oxygen

It appears that the model of eq. (2.7) can be also applied to the description of the dependence of the CBF on arterial partial pressure of oxygen $P_{O_2}$. For the data of (McDowall, 1966, Harper, 1989) (for dogs) a fit was obtained which is fully depicted in fig. 5.1. Above arterial pressure of about 50 mm Hg there is practically no change in CBF.

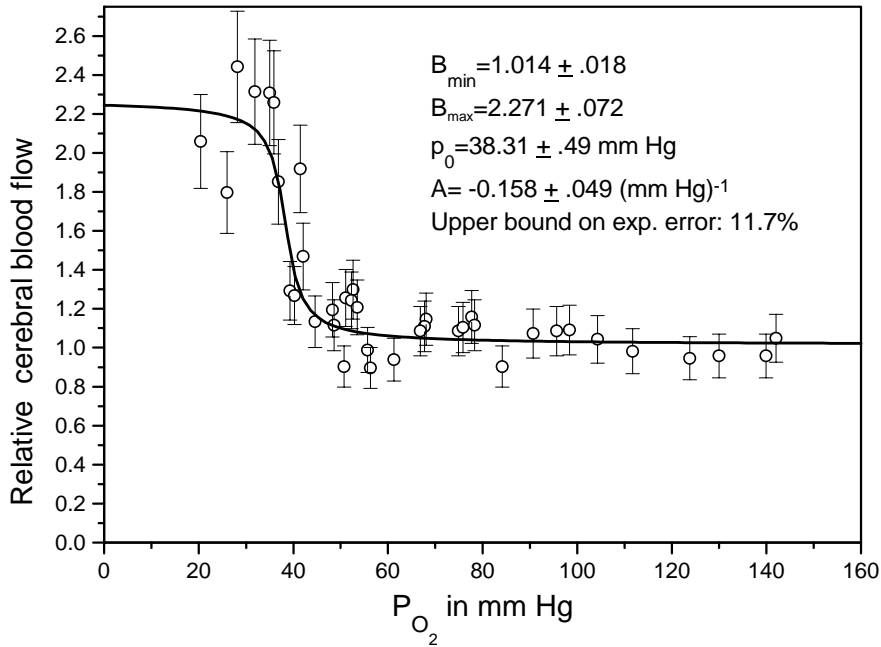

*Figure 5.1* The dependence of CBF of dogs on the partial tension of $O_2$, relative to its value above 70 mm Hg. The data are based on (McDowall,1966, Harper, 1989). The error
bars were added to the data according to the procedure outlined in sec. 3.

## 6. Discussion and conclusions

The major factors controlling the cerebral blood flow (CBF) are cerebral perfusion pressure (CPP), arterial partial pressure of oxygen ($P_{aO_2}$), cerebral metabolism, arterial partial pressure of carbon dioxide ($P_{aCO_2}$), and cardiac output (CO). The effect of $P_{aCO_2}$ is peculiar in being independent of autoregulatory CBF mechanisms and allows to explore the full range of the CBF. In Sec. 2 a simple model was derived describing the dependence of the CBF on $P_{aCO_2}$, and a simple formula, eq. (2.7), was derived. The model parameters $B_{max}$, $B_{min}$, A and $p_0$, have a simple meaning and can be determined easily from the experimental data. Although it appears that the minimal CBF $B_{min}$ reaches this value for negative $P_{a_{CO_2}}$, far away from the physical region, it strongly affects the near zero tensions. The other parameters: $B_{max}$, A and $p_0$ are more directly related to the data. $B_{max}$ can be associated with the dilation of the blood vessels and with the maximal CBF, therefore it is a good indicator of the CBF capacitance. We can expect that it should be age dependent and decline with age. The parameter A, the slope, can be connected with the adaptability to changes in $P_{a_{CO_2}}$. It should also decline with age. In this way we not only found an accurate procedure, but also found age dependent parameters related to the elasticity and adaptability of the blood vessels and CBF.

By analyzing the cardiovascular and the respiratory systems one can try to device exercises with the aim to improve the values of the parameters. In parallel, experiments are needed to study the age dependence of these parameters.

As the values of $p \equiv Pa_{CO_2}$ influence strongly the CBF, one can consider breathing exercises as means of increasing the elasticity of brain vessels. For example by alternately hyperventilating and hypoventilating (or breath holding) one can alternately shrink and dilate the cerebral blood vessels, exercising in this way their elasticity.

Theoretically one can with the increase of $Pa_{CO_2}$ increase to large extent the CBF, but in practice this has its dangers and also is very difficult to achieve, as with a slight change of $Pa_{CO_2}$ from normal, there is a strong urge to ventilate. For very advanced yogis the training of breath holding is of fundamental importance (Bernard, 1960, Rama 1986, 1988, Yeshe, 1998). The minimal time, in their opinion, to get real benefits, is 3 minutes, which is extremely difficult and dangerous for Westerners. But one should take into account that the aims of yogis are quite peculiar and that Westerners can benefit from much milder forms of exercise (Fried and Grimaldi, 1993). Indeed, the milder forms of pranayama (the yoga system of breathing) seems to be quite beneficial, as research around the world indicates (Chandra, 1994, Fisher, 1971, Juan et al., 1984, Kuvalayananda, 1933, Kuvalayananda and Karambelkar, 1957, Nagarathna, 1985, Patel, 1975, Stanescu et a., 1981, Rama et. al., 1979). In Russia the breathing exercises of K.P. Buteyko MD are well known (Buteyko 1983), but they are not documented in advanced scientific journals (outside Russia see Berlowitz, 1995, Hale, 1999). In this method the patients are learned to breathe superficially and to hold (out) the breath for about 1 minute. All this in order to increase the $Pa_{CO_2}$.

It seems to us that with rhythmic breathing exercises, with some mild breath holding, one can gradually build up the $Pa_{CO_2}$ towards beneficial values, including an improved mental activity. Although a large amount of information has been already gained (Kety and Schmidt, 1946, Kety and Schmidt, 1948, Sokoloff et al., 1955, Lassen, 1959, Sokoloff, 1960, Harper and Bell, 1963, Reivich, 1964, Ingvar and Risberg, 1965, Shapiro et al., 1965, Shapiro et al., 1966, Huber and Handa, 1967, Waltz, 1970, Fujishima et al., 1971, Harper et al., 1972, Kuschinsky et al., 1982, Paulson et al., 1972, Smith and Wollman, 1972, Symon et al., 1973, Ingvar and Schwartz, 1974, Fitch et al., 1975, Strandgaard et al., 1975, Jones et al., 1976, McKenzie et al., 1979a, 1979b,Yamaguchi et al., 1979, Maximilian et al., 1980, Yamamoto et al., 1980, Sokoloff, 1981, Gross et al., 1981, Kuschinsky et al., 1981, Gur et al., 1982, and others quated in this paper), more theoretical and experimental work is needed to clarify this issue.

The autoregulatory mechanisms of the brain, independent of $Pa_{CO_2}$, keeps the CBF constant as long as the CPP is below 160 mm Hg (and above 60 mmHg). With higher perfusion pressures the CBF and brain's temperature can be increased. Therefore the increased cardiac output (as evaluated in the Appendix A) can affect the brain if at the same time CPP exceeds 160 mm Hg. We expect that the increase of CBF will be limited by $B_{max}$ as this parameter is determined by the mechanical properties of blood vessels and the cranial space limits.

# Appendix A
## The effect of exercise on cardiac output and CBF

The highest efficiency for conversion of food energy into muscle work is at best 25%, the rest is converted into heat (Guyton, 1991). On the other hand the amount of heat produced in the body is directly proportional to the oxygen consumption. At rest the rate of consumption is about 0.2 to 0.3 L/min, and it can increase to 3-6 L/min during maximal exercise Guyton, 1991), depending, among other things, upon age, sex and level of fitness. The energy production is about 5 kcal per 1 L of oxygen consumed. For example, in running, the energy production is approximately 0.2 mL of oxygen per 1 kg of body weight and per meter run. Accordingly, a 70 kg runner, running 2000 m, will produce about 140 kcal of energy. If he will be well insulated with a proper dressing, great part of this energy may be used to increase his body temperature, and for releasing humidity from the lungs. Remembering that 15% of the cardiac output goes to the brain, we may infer that the brain can receive a large amount of heat energy.

According to Hodgkin and Huxley the reaction rate R in the axon changes with the temperature change $\Delta T$ in the following way (Hobbie, 1988): $R = 3^{\Delta T/10^{o}C}$.

The cardiac output C is directly proportional to the work output W and to the oxygen consumption $\dot{V}_{O_2}$. From (Guyton, 1991), which considers typical experimental results, we derived the following linear relations:

C=6.8 +7($\dot{V}_{O_2}$ - 0.25 L/min) L/min ,   (A.1)

C=[6.8+0.0141·W] L/min,   (W in kg·m/min) ,   (A.2)

$\dot{V}_{O_2}$ =[0.25 +0.00202·W] L/min, (W in kg·m/min) .   (A.3)

Let us take as an example a runner running 2000m and consuming (additional to the con- sumption at rest of about .25L/min) 28 L of oxygen. His cardiac output will depend on the runner's speed. For example if he will cover the 2km distance in 14 min. (a mediocre time), we will find from eq. (A.1) that his cardiac output will be 20.8 L/min, i.e. 3 times larger than his cardiac output at rest. We can see that even a moderate exercise can in- crease to a large extent the cardiac output. If the body will be well insulated, a large amount of energy (in the form of heat) could have been transmitted to the brain.

The autoregulatory mechanisms of the brain forbid it from occurring as long as the CPP is below 160 mm Hg. With higher perfusion pressures the CBF and brain's temperature can be increased, but the CBF should be always limited by the value $B_{max}$ which was found in Sec. 2.

According to (McArdle et al., 1996) the CBF increases during exercise by approximately 25 to 30% compared to the flow at rest (Herlhoz et al., 1987, Thomas et al., 1989).